# Cometary glycolaldehyde as a source of pre-RNA molecules

Nicolle E. B. Zellner[1], Vanessa P. McCaffrey[2], Jayden H. E. Butler[1,3]

[1]Department of Physics, Albion College, Albion, Michigan USA 49224

[2]Department of Chemistry, Albion College, Albion, Michigan, USA 49224

[3]Department of Physics, California State University – Los Angeles, Los Angeles, CA, USA 90032

Corresponding Authors:

Nicolle Zellner
611 E. Porter St.
Physics, Albion College
Albion, MI 49224
517-629-0465

nzellner@albion.edu

Vanessa McCaffrey
611 E. Porter St.
Chemistry, Albion College
Albion, MI 49224
517-629-0622

vmccaffrey@albion.edu

**Abstract**

Over 200 molecules have been detected in multiple extraterrestrial environments, including glycolaldehyde ($C_2(H_2O)_2$, GLA), a two-carbon sugar precursor that has been detected in regions of the interstellar medium. Its recent *in situ* detection on the nucleus of comet 67P/Churyumov–Gerasimenko and via remote observations in the comae of others, provides tantalizing evidence that it is common on most (if not all) comets. Impact experiments conducted at the Experimental Impact Laboratory at NASA's Johnson Space Center have shown that samples of GLA and GLA mixed with montmorillonite clays can survive impact delivery in the pressure range of 4.5 GPa to 25 GPa. Extrapolated to amounts of GLA observed on individual comets and assuming a monotonic impact rate in the first billion years of solar system history, these experimental results show that up to $10^{23}$ kg of cometary GLA could have survived impact delivery, with substantial amounts of threose, erythrose, glycolic acid, and ethylene glycol also produced or delivered. Importantly, independent of the profile of the impact flux in the early solar system, comet delivery of GLA would have provided (and may continue to provide) a reservoir of starting material for the formose reaction (to form ribose) and the Strecker reaction (to form amino acids). Thus, comets may have been important delivery vehicles for starting molecules necessary for life as we know it.

1. Introduction

Over its 4.5-billion-year history, the planets and moons of the solar system have been battered by comets, asteroids, and meteorites, which likely delivered organic molecules to their surfaces, including Earth's. The rate of this delivery and the subsequent biochemistry of the surviving molecules, however, are the subjects of much debate (e.g., Chyba *et al.* 1990; Owen 1998; Pasek and Lauretta, 2008; Zellner 2017). Estimates of the current flux of meteoroids at Earth is ~$10^7$ kg/yr (e.g., Love and Brownlee 1993; Pokorný *et al.* 2019), and the flux of extraterrestrial material was certainly higher in the early solar system. Based on the abundance of highly siderophile elements on the Moon, Morbidelli *et al.* (2018) reported that up to $10^{-3}$ Earth masses ($10^{21}$ kg) of material was delivered to Earth in its first billion years, and Chyba *et al.* (1990) calculated that organic molecules were being delivered at a rate of $10^{13}$ kg/year during the same time period. Whittet (2017) suggested that, based on water delivered by asteroids, the accompanying mass of organic matter useful to pre-biotic chemistry would be ~$10^{19}$ kg; comets may have delivered 10-20% of that amount (e.g., Marty and Meibom, 2007; Barnes *et al.* 2016; Marty *et al.* 2016; Avice *et al.* 2017; Nesvorney *et al.* 2017; Morbidelli *et al.* 2018). These extraterrestrial organic materials could have been significant sources of starting materials for the formation of simple biomolecules, leading to life on Earth (e.g., Anders 1989; Chyba *et al.* 1990; Ehrenfreund and Charnley 2000). For example, extraterrestrial amino acids (Pizzarello *et al.* 1991; Botta and Bada 2002; Elsila *et al.* 2009; Burton *et al.* 2011; Cody *et al.* 2011; Glavin *et al.* 2012), sugar derivatives (Cooper *et al.* 2001; Cooper and Rio 2016), and sugar (Furukawa *et al.* 2019) may have provided important materials for the formation of DNA and RNA, the genetic components of life as we know it.

With over 200 molecules detected in multiple different extraterrestrial environments (e.g., Smith 2012; McGuire 2018), studies of their existence in interstellar ice and grain mantles (e.g., Charnley *et al.* 2001; Dworkin *et al.* 2001; Bernstein *et al.* 2002; Nuevo *et al.* 2010), comets (Elsila *et al.* 2009), and meteorites (e.g., Pizzarello *et al.* 1991; Cooper *et al.* 2001; Botta and Bada 2002; Burton *et al.* 2011; Glavin *et al.* 2012; Cooper and Rio, 2016) have become commonplace. Additionally, impact experiments conducted on many of these biologically relevant molecules show that conditions under which most impacts occur do not totally destroy them. In fact, amino acids and simple sugars that were subjected to pressures of 5 to 25 GPa (e.g., Peterson *et al.* 1997; Blank *et al.* 2001; Bertrand *et al.* 2009; Martins *et al.* 2013; McCaffrey *et al.* 2014) demonstrated that not only will a portion of the molecules remain intact, but that the molecule can also undergo a series of polymerization and/or racemization reactions. Thus, both the survival of these molecules and the subsequent production of other molecules via shock chemistry at the time of impact indicate that comets and asteroids, along with meteorites and interplanetary dust particles, may have been very effective delivery vehicles for biomolecules.

**1.1 Sugar Molecules in Extraterrestrial Materials**

*1.1.1 Observations in Star-forming Regions*

Glycolaldehyde ($C_2(H_2O)_2$, GLA), a two-carbon sugar precursor, is ubiquitous in space and has been proposed to form on grain surfaces (Woods *et al.* 2012) or inside icy grain mantles that are photo-processed by ultraviolet starlight (Sorrell 2001; Bennett and Kaiser 2007; Jalbout *et al.* 2007). It has been detected in the molecular cloud Sagittarius B2 (e.g., Hollis *et al.* 2000, 2001, 2004; Halfen *et al.* 2006; Li *et al.* 2017), the hot molecular core G31.41+0.31 (Beltrán *et al.* 2009), the protostar IRAS 16293-2422 (Jørgensen *et al.* 2012), and in the Orion-KL star-forming region (Pagani *et al.* 2017). GLA is of considerable interest to astrobiology and has been postulated to be a precursor in the synthesis of ribose in interstellar space (Jalbout *et al.* 2007) via the formose reaction (Lambert *et al.* 2010; Kim *et al.* 2011).

Ethylene glycol ($HOCH_2CH_2OH$; EG) is also commonly found in space and in many of the same locations as GLA (e.g., Hollis *et al.* 2002; Li *et al.* 2017; Pagani *et al.* 2017). EG is the first of the series of polyols and while not of direct biological importance, its detection in these regions suggests that it could be an important source of reactive carbon (e.g., Brown *et al.* 1986).

*1.1.2 Detections on Comets*

The detection of GLA around the protostar IRAS 16293-2422 (Jørgensen *et al.* 2012) and in the Orion-KL star-forming region (Pagani *et al.* 2017) provided tentative evidence that this molecule could be easily incorporated into the debris disk and/or planetestimals orbiting around young stars. Models simulating solar system formation show that incorporation and preservation of biomolecules in planetesimals (e.g., Marboeuf *et al.* 2014; Hallis *et al.* 2015) is possible, so it is not unreasonable to suggest that the presence of GLA and EG on comets and asteroids could be ubiquitous. Indeed, Crovisier *et al.* (2004) announced the first radio detections of EG in the coma of Comet C/1995 O1 (Hale-Bopp), with detections of these same molecules in the comae of C/2012 F6 (Lemmon; Biver *et al.* 2014), C/2013 R1 (Lovejoy; Biver *et al.* 2014), and C/2014 Q2 (Lovejoy; Biver *et al.* 2015) subsequently announced. In 2015, Goesmann *et al.* reported *in situ* detections of both GLA and EG on the nucleus of comet 67P/Churyumov–Gerasimenko (hereafter, 67P).

Aldehydes and their reduced alcohols (GLA and EG in this work) are commonly found together in the ISM and in cometary comae. The formation of EG and GLA on comet nuclei has been postulated to proceed via radical mechanisms promoted by vacuum UV irradiation. Both experimental (Butscher *et al.* 2016) and theoretical results (Wang and Bowie, 2010) have shown the plausibility of these reaction pathways. In laboratory irradiation experiments, the EG/GLA ratio was measured to be 5/1 (Crovisier *et al.* 2004), similar to the ratios reported for many comets

(Coutens *et al.* 2015). This large difference has been explained in two ways. First, the predominance and ease of the radical dimerization reaction of •CH$_2$OH to form EG, compared to the heterogeneous reaction of •CH$_2$OH and •C(O)H to form GLA; and second that GLA is inherently more reactive than EG and has a much shorter half-life on the comet body (Butscher *et al.* 2015).

These comets on which GLA and EG have been detected are classified as nearly isotropic comets (NICs) that originated in the Oort Cloud (Lemmon, Lovejoy 2013, Lovejoy 2014, Hale-Bopp) and Jupiter Family Comets (JFCs; 67P) that originated in the Kuiper Belt, implying that GLA and EG might be found on every kind of comet in our solar system. Thus, if even just 10% of impactors in the early solar system were comets (Marty and Meibom, 2007; Marty *et al.* 2016; Avice *et al.* 2017), GLA and EG would have been delivered to young planets and moons in large amounts.

*1.1.3 Detections in Meteorites*

In 2001, Cooper and coworkers identified dihydroxyacetone (C$_3$H$_6$O$_3$, the simplest ketose) in samples of the Murchison and Murray meteorites, both CM2 carbonaceous chondrites with extensive aqueous alteration. The extraterrestrial origins of the compound were confirmed through isotopic analysis (Cooper and Rios, 2016). Many polyol sugar derivatives were also identified in this sample, including C3-C6 sugar alcohols, C3-C6 sugar acids and numerous C4-C6 deoxysugar acids (Cooper *et al.* 2001). At the time, no aldoses, including GLA, had been found in either the Murchison or the Murray samples, or in any other meteorite sample that could be conclusively identified as being extraterrestrial in origin, likely due to its high reactivity (*vida supra*). However, ethylene glycol (the reduced variant of GLA) has been found in several meteorite samples in very high concentrations relative to other organics (Cooper and Rios, 2016), leading to some speculation that it could have been synthesized from GLA during alterations of the parent body (Feedosev *et al.* 2015 and references therein).

The first conclusive identification of ribose and several other aldoses was in 2019 by Furukawa *et al*. Three different carbonaceous chondrite meteorite samples, including Murchison, were determined to contain sub-0.5 to 150 ppb concentrations of ribose and the three additional possible aldopentoses. The authors speculated that these pentoses could have been formed through the formose reaction of reactive aldehydes present in the parent body. A control reaction of GLA and formaldehyde replicated the relative concentrations of the sugars, lending support to the idea that these compounds could be formed through the formose reaction of GLA with *in situ* formaldehyde.

**1.2 Experimental Production of Biomolecules**

The impact production of biomolecules has been demonstrated using a variety of experimental apparatuses, including vertical guns (e.g., Zellner *et al.* 2012), flat-plate accelerators (e.g., Bertrand *et al.* 2009; McCaffrey *et al.* 2014), light-gas guns (e.g., Blank *et al.* 2001; Bowden *et al.* 2008; Martins *et al.* 2013), and lasers (e.g., Scattergood *et al.* 1989; Meinert *et al.* 2016). These experiments can, to a good approximation, simulate what we know of impact conditions on planetary bodies (e.g., Holsapple 1993; DeCarli *et al.* 2004). Thus, the experiments can provide a baseline for models that aim to understand how biologically relevant molecules are both delivered and affected by impacts (e.g., Steel 1998; Pierazzo and Chyba 1999, 2006; Ross 2006). In all experiments, the impacts affected the chemistry of these biomolecules, producing both simple and complex species. While some degradation of the starting molecule was observed, the overall conclusion, as described in these studies, is that biomolecules delivered by comets and asteroids (and meteorites and interplanetary dust particles) are not necessarily adversely affected under most impact conditions.

As a direct application of how shock can affect the chemistry of GLA delivered by a comet, the dimerization reaction of GLA to produce the four-carbon sugars erythrose and threose was seen in the impact-shock experiments of GLA stabilized with montmorillonite clays (McCaffrey *et al.* 2014). McCaffrey *et al.* (2014) showed that up to 95% of GLA can survive low-pressure impacts; that EG (a diol that is the reduced form of GLA) can be produced; and that erythrose and threose can be formed. While EG decomposes quickly in the environment, erythrose and threose can persist under a variety of conditions, and the identification of these latter compounds through this relatively simple reaction has interesting implications for prebiotic chemistry. For example, in trying to understand and explain the chemical origins of ribonucleic acid (RNA), many researchers proposed that there must be a "simpler" genetic polymer that could store genetic information and would eventually evolve into RNA and then to DNA (e.g., Woese 1967; Crick 1968; Orgel 1968; Gilbert 1986). There have been several polymers postulated to fill in this evolutionary gap, including peptide nucleic acid (PNA, Nelson *et al.* 2000) and threose nucleic acid (TNA). The latter is especially relevant to this study as threose makes up its sugar backbone (e.g., Robertson and Joyce, 2012 and references therein). The building blocks of this proposed RNA-precursor should be easily synthesized under prebiotic conditions (Orgel 2000). In 2000, Eschenmoser showed that oligonucleotides based on TNA exhibit Watson-Crick base pairing that is stronger than in RNA (Schöning *et al.* 2000). Short repeats of these TNAs would also bind to complementary strands of RNA and DNA (Anasova *et al.* 2016), suggesting that information could be swapped between TNA and RNA. It has also been shown that threose can be stabilized with borate-containing minerals (Kim *et al.* 2011). This stabilization is necessary in order to preserve the carbohydrate long enough for it to react with the other components of the TNA backbone.

**1.3 Catalyzing Materials**

Montmorillonite clay, formed by the weathering of volcanic ash (Papike 1969), might have played an important role in the evolution of prebiotic molecules on an early Earth and other planetary bodies (Joshi *et al.* 2009; Delano *et al.* 2010; Jheeta and Joshi 2014). Montmorillonites are layered minerals with octahedral alumina ($AlO_6$) sheets interspersed between sheets of tetrahedral silica ($SiO_4$). The layered nature of these minerals gives them extremely high surface areas, increasing their ability to adsorb water and organics to their surface (Cleaves *et al.* 2012). The interesting catalytic properties of montmorillonite clays come from the rich substitution of iron(II), iron(III), magnesium(II) and other metal ions in the octahedral alumina layers (e.g., Ferris, 2005), and the ability of these clays to adsorb organic compounds has been well studied, including the effect of ion substitution on the absorption of a variety of functional groups (Hashizume *et al.* 2010). The majority of the binding is thought to occur between the sheets and the clays will expand to accommodate these compounds. However, it has also been shown that some organic compounds and water in particular can bind on the broken edges of the sheets to the aluminum(III) ions (Pedreira-Segade and Rogers, 2019). This coordination of water can change the local pH of the clays and can dramatically influence their catalytic properties (Ferris and Ertem, 1993).

Montmorillonite clays have been shown to catalyze the polymerization of activated nucleotides to form oligonucleotides that range in length from dimers to 50mers, depending on the structure of the nucleotides and the different modifications used to activate them (Ferris *et al.* 1996; Ferris 2002). The montmorillonite clays were also found to be slightly regio- and sequence-selective (Ferris and Miyakawa 2003), limiting the number of isomers that can be formed in a reaction, a selectivity that is important when considering their potential role in the evolution of biomolecules. Importantly, montmorillonite clays are abundant in the solar system; their ubiquitous presence on planetary objects, including those suspected of being habitable (e.g., Europa, Shirley *et al.* 2013; Mars, Bishop *et al.* 2008; Wray *et al.* 2008; Ehlmann *et al.* 2011; Sun and Milliken, 2014), on comets (e.g., Lisse *et al.* 2006), and in meteorites (e.g., Anders *et al.* 1973; MacKinnon 1985; Zolensky and Keller, 1991; Busek and Hua, 1993; Cody *et al.* 2011) suggests that the formation and preservation of biomolecules in the presence of montmorillonite clays may be common.

**1.4 Impact Flux in the Early Solar System**

The impact flux in the inner solar system, especially during the first 700 million to 1 billion years of Earth's history, has been debated since the arrival and study of Apollo lunar samples in 1969. Since then, development and utilization of high-precision instruments, improved sample analysis techniques, and high-resolution orbital data have allowed for a paradigm shift in the interpretation of the lunar impact flux (as described in, for example, Bottke and Norman 2017; Zellner 2017; Michael *et al.* 2018; Hartmann 2019). In particular, the "lunar cataclysm", a proposed "spike" in the lunar impact flux at ~3.9 billion years ago that was presumed to have formed all of the nearside basins (e.g., Mare Imbrium, Mare Serenitatis), has been called into question. Lunar sample evidence (i.e., composition and age; e.g., Grange *et al.* 2013; Liu *et al.* 2012; Merle *et al.* 2014; Mercer *et al.* 2015) indicates that ejecta from Mare Imbrium likely contaminated the nearside samples collected by the astronauts at all six Apollo landing sites. In turn, this sample bias likely affected the samples' age interpretations. Dynamical models that describe impact flux scenarios are being revised as new data are published and older data are re-analyzed and interpreted. For example, a prolonged impact flux and a longer decline over 400-600 million years after about 4.1 billion years ago has been proposed to better reflect the lunar impact sample and terrestrial impact data (e.g., Bottke *et al.* 2012; Morbidelli *et al.* 2012, 2018). Additionally, an impact flux that declines monotonically has been reconsidered to explain the abundance of highly siderophile elements on the Moon (Morbidelli *et al.* 2018; Zhu *et al.* 2019), metamorphic histories and ages of martian zircon and baddeleyite grains (Moser *et al.* 2019), and martian cratering chronology (Werner, 2019). This monotonic flux of impactors would have delivered $10^{11}$ and $10^{12}$ kg of extraterrestrial material to Mars and Earth, respectively, each year (Morbidelli *et al.* 2018).

The relative contributions of biogenic material from comets and asteroids are not well-constrained and the timing of impact delivery depends on factors (e.g., timing, duration, magnitude) related to planetary migration or other dynamical instabilities in the solar system's disk. Crater sizes, shapes, and distributions on the lunar surface have been interpreted to suggest one or two (or more) impacting populations struck the Moon during the early bombardment era in the solar system (e.g., Gomes *et al.* 2005; Fassett and Minton, 2013), and chemical signatures in lunar Apollo and meteorite samples (e.g., Kring and Cohen 2002; Morbidelli *et al.* 2018) indicate asteroid impacts were prevalent at this time. Suggestions that 10% (e.g., Marty and Meibom, 2007;

Marty *et al*. 2016; Avice *et al*. 2017) to 20% (Barnes *et al.* 2016) of impactors were comets have been proposed, but the timing of when comets were delivered remains uncertain (e.g., Zahnle *et al*. 2003; Jørgensen *et al.* 2009; Nesvorný *et al*. 2017; Rickman *et al*. 2017; Morbidelli *et al*. 2018). Additionally, an understanding of the specific characteristics and compositions of comets continues to evolve as more are discovered and observed, causing challenges in categorizing them into specific taxonomies (Mumma and Charnley, 2011).

Nonetheless, as noted above, these impactors would have delivered large amounts of organic material, including water, amino acids, and sugar molecules, to Earth and other bodies in the inner solar system. It is the specific delivery, survival, and polymerization of the sugar pre-cursor molecule GLA that we describe herein. Its detection on different kinds of comets, the historical impact delivery of comets to habitable moons and planets, and the importance of GLA as a starting molecule in the formose and Strecker reactions prompted an investigation as to how much GLA (and subsequent polymerized molecules) would be available as drivers of pre-biotic chemistry.

## 2. Materials and Methods

### 2.1 Impact Survival of GLA

To understand the survivability and reactivity of GLA under impact conditions, McCaffrey *et al.* (2014) conducted experiments at a variety of impact pressures, with both neat GLA and GLA mixed in a 1:20 ratio with montmorillonite clay. As reported therein, GLA can survive and react under conditions that mimic meteoritic impact events with the surface of the Earth when stabilized with a mineral matrix; results are summarized in Table 1. When the impact pressures are low (simulating an impact at low-inclination angle), there is survivability of 95% of the original amount of GLA. Comparison of the samples before and after impact reveals that there is little change in the color or texture of the GLA/clay mixtures.

However, as the pressure is increased to ~25 GPa, which is more typical of a vertical impact, the survivability of the GLA drops dramatically. The samples after the impact event are dark brown in color (compared to a sandy white color before impact) and the GLA/clay mixtures are brittle and compacted. Around 5% of the original organic matter was recovered from the GLA/clay mixtures at these pressures.

### 2.2 Identification of Tetroses, Ethylene Glycol and other Polyhydroxylated Compounds

In addition to the recovered GLA, all impact samples contained small amounts of four-carbon sugars. The sugars produced were identified through retention time and mass spectrum matching to be threose and erythrose in an approximately 6:1 ratio, respectively. The amount of these carbohydrates is enhanced relative to the amounts found in control samples that were not subjected to the impact experiments. Additional polyhydroxylated compounds were seen in the samples, including glycolic acid and glycerol (4.65 GPa and 12 GPa only), several unidentified polyols, and ethylene glycol (25.1 GPa only). Details of the recovery and analyses are described in McCaffrey *et al.* (2014); recovery percentages from that study are shown in Table 1.

**Table 1.** Results of impact experiments showing survival of GLA and production of threose, erythrose, ethylene glycol (EG), glycolic acid, and glycerol under a variety of impact pressures. ND = not detected. Data are from McCaffrey *et al.* (2014).

|  | **Unshocked GLA/Clay (1:20)** | **4.65 GPa** | **12 GPa** | **25.1 GPa** |
|---|---|---|---|---|
| **Percent Recovery of GLA** | -- | 95% | 96% | 6.5% |
| **Threose (all forms)** | 0.4% | 3.6% | 1.2% | 1.2% |
| **Erythrose** | 0.2% | 0.6% | 0.3% | 0.2% |
| **Ethylene Glycol** | ND | ND | ND | 0.5% |
| **Glycolic Acid** | 0.62% | 1.17% | 0.62% | 0.01% |
| **Glycerol** | 0.16% | 0.57% | 0.30% | 0.15% |

## 2.3 Calculating Delivery Amounts of Sugar Molecules

In order to estimate the amount of GLA, EG, erythrose, and threose that would have been delivered intact and/or produced by comet impact, we used published information about the sizes of nuclei for comets at which GLA and/or EG had been observed and the reported observed amount of each (relative to water). Except in the case of comet 67P, the size and density of the nucleus for each of the comets with detected amounts of GLA and EG in the comae are not well-constrained. However, using a range of published values, a size can be estimated. Additionally, assuming a spherical nucleus and uniform (homogeneous) distribution of material (Preusker *et al.* 2015; Pätzold *et al.* 2016; Herny *et al.* 2018), as well as the understanding that organic molecules detected in comet comae must be present in the nuclear ice (e.g., Charnley *et al.* 2002; Rodgers and Charnley, 2002), the amount of GLA and EG in each comet can be reasonably estimated. Other uncertainties arise from estimates of the amount of water in and the particular density of each comet. "Dirty snowballs" imply that comets are mostly water-ice, and compositions ranging from 30% (Greenberg 1998) to 50% (Ehrenfreund *et al.* 2004) water/water-ice have been reported for comets in general; a value of 30% is used here. Similarly, comet densities can range from 0.2 to 1.65 g/cm$^3$ (Greenberg, 1998; Davidsson and Gutierez, 2004; A'Hearn *et al.* 2005) and a value of 0.6 g/cm$^3$ (Britt *et al.* 2006) is used here.

Considering these reasonable parameters, amounts of GLA that would exist in each comet were calculated by estimating the mass of water present on each comet, assuming that the composition of the comet is 30% water-ice (except in the case of comet 67P, where a published value of 15% was used; Fulle *et al.* 2016). The reported mole fractions of GLA and EG for each comet were converted to mass fractions (see Table 2 for the values and references) and from this, a reasonable amount of GLA and EG was determined to be present on each of the comets before impact. Then, using the experimental results of McCaffrey *et al.* (2014), amounts of GLA that would survive comet delivery and amounts of EG, erythrose, and threose that would be produced when the comet impacted a surface were derived. While not determined directly, the survivability of EG was assumed to be the same as that of GLA under similar conditions. Values in Table 3 were determined by multiplying the calculated amounts of GLA and EG present on the comet by reactivities of GLA, glycerol, threose, erythrose and glycolic acid under impact conditions (Table

1). Finally, estimates of the total delivered amounts of GLA, EG, erythrose, and threose available on early Earth are reported in Table 4, assuming a monotonic decline in impact flux in the first billion years of solar system history.

## 3. Results

For Hale-Bopp, Lemmon, Lovejoy 2013, and Lovejoy 2014, the observations of GLA and EG in the comae were reported relative to water, and the published water production rates were used to estimate the masses of each molecule observed. Comet 67P benefitted from *in situ* measurements of GLA and EG. Total comet masses were estimated using published diameters and densities when known; physical parameters for the five comets are shown in Table 2.

Table 3 shows the amounts of GLA that would survive impact at 4.6 GPa (velocity ~ 11.2 km/sec, impact angle ~ 7.7° from the horizontal; Blank and Miller, 1998; Blank *et al*. 2001) and 25 GPa (11.2 km/sec, impact angle >26.5° from the horizontal; Blank and Miller 1998; Blank *et al*. 2001), along with the amounts of threose, erythrose, glycerol, and glycolic acid that would have been produced during impact of each comet. Resulting yields for each impact pressure scenario (Table 1; McCaffrey *et al*. 2014) have been applied to total amounts reported in Table 2. Additional data resulting from different initial starting parameters (e.g., different size nucleus) can be found in the supplementary section of the published journal article.

The estimated minimum and maximum amounts of GLA, threose, erythrose, glycerol, glycolic acid, and ethylene glycol that would have been delivered or produced by comet impacts during the first billion years of Earth's history are shown in Table 4, assuming the higher shock pressure more typical of comet velocities (e.g., Blank *et al.* 2001). As the presumed smallest comet in this study, compositions and characteristics of Lovejoy 2014 are used to derive the minimum delivered or produced masses of the sugar molecules (up to $10^{16}$ kg, in the case of GLA), assuming all of the impacting comets have the characteristics and composition of Lovejoy 2014. As the largest comet, those of Hale-Bopp are used to derive the maximum masses (up to $10^{23}$ kg, in the case of GLA), assuming all of the impacting comets have the characteristics and composition of Hale-Bopp. Calculated masses of GLA and EG on 67P benefitted from *in situ* measurements, so maximum masses delivered by comets similar to 67P are also reported (up to $10^{21}$ kg, in the case of GLA). The amounts were calculated assuming that comets made up 10% of the impactors contributing to the $10^{12}$ kg/yr (Morbidelli *et al.* 2018) to $10^{13}$ kg/year (Chyba *et al.* 1990) of material delivered to an early Earth and considering the fraction of GLA per comet (Table 2) as well as its survivability (Table 3).

**Table 2.** Physical parameters for the five comets reported herein. A density of 0.6 g/cm$^3$, the reported minimum radius, and a spherical shape were used to calculate the masses of the NICs. A homogeneous distribution of molecules in the nucleus was used to determine the mass of GLA in each comet. NIC = Nearly Isotropic Comet, JFC = Jupiter Family Comet. Mole % of GLA relative to water assumes 30% water-ice in the comet (Greenberg 1998). GLA data for the NICs are from Crovisier *et al.* (2004) and Biver *et al.* (2014, 2015), and GLA and other data for 67P, which has two lobes (of radii 2.05 km and 2.15 km), are from Goesmann *et al.* (2015), Preusker *et al.* (2015), Fulle *et al.* (2016), and Pätzold *et al.* (2016).

| Comet Name | Diameter (km) | Mass ($\times 10^9$ kg) | Mole% of GLA Relative to Water | Mass of GLA ($\times 10^6$ kg) | %GLA per comet | Mole% of EG Relative to Water | Mass of EG ($\times 10^6$ kg) | %EG per comet |
|---|---|---|---|---|---|---|---|---|
| **Hale-Bopp** (NIC) | 13.5 | $6.2 \times 10^6$ | 0.04 | $2.5 \times 10^6$ | 0.1 | 0.25 | $1.6 \times 10^7$ | 1 |
| **Lemmon** (NIC) | 1.0 | $2.5 \times 10^3$ | 0.08 | 2010 | 0.2 | 0.24 | 6231 | 0.5 |
| **Lovejoy 2013** (NIC) | 1.0 | $2.5 \times 10^3$ | 0.07 | 1759 | 0.2 | 0.35 | 9087 | 1 |
| **Lovejoy 2014** (NIC) | 0.25 | 26 | 0.016 | 4.2 | 0.04 | 0.07 | 19 | 0.2 |
| **67P** (JFC) | 4.2 | $10^4$ | 0.4 | $4.2 \times 10^4$ | 1.1 | 0.2 | $1.0 \times 10^4$ | 0.6 |

**Table 3.** Estimated minimum masses of sugar molecules that would survive or be produced during impact, at the given impact pressures, using physical parameters (Table 2) and applying resulting yields (Table 1). NIC = Nearly Isotropic Comet, JFC = Jupiter Family Comet. (Updated from Zellner *et al.* 2018.)

| Comet Name (minimum diameter, family) | Impact Pressure (GPa) | Impact-Surviving GLA ($\times 10^6$ kg) | Impact-Produced Threose ($\times 10^6$ kg) | Impact-Produced Erythrose ($\times 10^6$ kg) | Impact-Produced Glycerol ($\times 10^6$ kg) | Impact-Produced Glycolic Acid ($\times 10^6$ kg) | Impact-Surviving Ethylene Glycol ($\times 10^6$ kg) |
|---|---|---|---|---|---|---|---|
| **Hale-Bopp** (13.5 km, NIC) | 4.6 | $2.4 \times 10^6$ | $8.1 \times 10^4$ | $1.4 \times 10^4$ | $1.3 \times 10^4$ | $2.9 \times 10^4$ | $1.5 \times 10^7$ |
| | 25 | $1.6 \times 10^5$ | $2.8 \times 10^4$ | 3708 | 791 | 247 | $1.0 \times 10^6$ |
| **Lemmon** (1.0 km, NIC) | 4.6 | 1909 | 66 | 11 | 11 | 24 | 5920 |
| | 25 | 131 | 23 | 3 | 0.6 | 0.2 | 405 |
| **Lovejoy 2013** (1.0 km, NIC) | 4.6 | 1671 | 58 | 10 | 10 | 21 | 8633 |
| | 25 | 114 | 20 | 3 | 0.6 | 0.2 | 591 |
| **Lovejoy 2014** (0.25 km, NIC) | 4.6 | 4 | 0.1 | 0.02 | 0.02 | 0.05 | 18 |
| | 25 | 0.3 | 0.05 | 0.01 | 0.0013 | 0.0004 | 1.23 |
| **67P** (4 km, JFC) | 4.6 | $4.0 \times 10^4$ | 1392 | 241 | 229 | 495 | $2.0 \times 10^4$ |
| | 25 | 2750 | 486 | 63 | 14 | 4 | 1400 |

**Table 4.** Estimated minimum and maximum masses of sugar molecules delivered or produced by comet impacts at a pressure of 25 GPa during the first billion years of Earth's history. Total delivered material assumes comets comprised 10% of impactors at a rate of $10^{12}$ kg/yr (minimum; Morbidelli et al. 2018) to $10^{13}$ kg/yr (maximum; Chyba et al. 1990).

|  | GLA (kg) | Threose (kg) | Erythrose (kg) | Glycerol (kg) | Glycolic Acid (kg) | Ethylene Glycol (kg) |
|---|---|---|---|---|---|---|
| **Minimum** | $3.0 \times 10^{16}$ | $5.0 \times 10^{15}$ | $1.0 \times 10^{15}$ | $1.3 \times 10^{14}$ | $4.0 \times 10^{13}$ | $1.2 \times 10^{17}$ |
| **Maximum** | $1.6 \times 10^{23}$ | $2.8 \times 10^{22}$ | $3.7 \times 10^{21}$ | $7.9 \times 10^{20}$ | $2.5 \times 10^{20}$ | $1.0 \times 10^{24}$ |
| **67P (Maximum)** | $2.8 \times 10^{21}$ | $4.9 \times 10^{20}$ | $6.3 \times 10^{19}$ | $1.4 \times 10^{19}$ | $4.0 \times 10^{18}$ | $1.4 \times 10^{21}$ |

## 4. Discussion

Swarms of comets and asteroids were likely present during the final accretionary and sweep-up phases in the early solar system and impacts were probable. Calculations based on experimental results, telescope observations, and/or spacecraft instrument measurements of cometary GLA in NICs and JFCs (Table 3, Table 4) show that large amounts of impactor GLA and impact-derived sugar molecules would have been abundant on planetary bodies when the early solar system was experiencing much more frequent bombardment. In the most generous case reported in this study (13-km Hale-Bopp impacting at 25 GPa; Table 4), masses of molecules from ~0.01 to ~100× the mass of the Moon ($10^{22}$ kg) would have been available for pre-biotic chemistry. *In-situ* measurements of GLA on 67P allowed estimates to be more refined: up to $10^{21}$ kg (10% the mass of the Moon) of GLA would have survived impact.

Atmospheric entry of bolides, however, can have deleterious effects on bolide size and composition. Hills and Goda (1993) demonstrated that large objects ablate as they decelerate through the atmosphere, and Mehta et al. (2018) reported that ~60 to >99% of the object's mass could be lost to ablation. Even so, because comets dissipate energy at high altitude, the blast wave is weaker (Hills and Goda, 1993), which may allow for the survival of organics. Moreover, any material affected by the blast wave does not necessarily disappear - it can be transformed into other material, as demonstrated by results of impact experiments and analyses that showed survival and production of shocked organic molecules (e.g., Petterson et al. 1997; Blank et al. 2000; Court and Sephton, 2009; McCaffrey et al. 2014). The blast wave can also carry off excess heat, preserving interior organic materials, as observed in extraterrestrial samples of all sizes (e.g., Murchison, Pizarello et al. 1991; Tagish Lake, Brown et al. 2000; micrometeorites, Glavin and Bada, 2001). Additional opportunities for survival of organics include lower shock pressures experienced during impact into a porous target (e.g., Love et al. 1993; Wünnemann et al. 2005); the decay of peak shock pressure during impact into water (e.g., Artemieva and Shuvalov, 2002; Artemieva and Pierazzo, 2011); and smaller impacts that experience only short-lived thermal anomalies (e.g., Losiak et al. 2020). Fragmentation can additionally allow survival of organic materials because it enables a deceleration in velocity (e.g., Artemieva and Pierazzo, 2009) and allows ablation to act over a much larger surface area (Mehta et al. 2018). Though large objects may be less effective "organic point sources" because their fragments may be spread out over large areas (Mehta et al.

2018), we argue that in any case, material is still relatively locally concentrated. Organic material need only be concentrated under the "right" conditions for chemical reactions to occur.

Impacts by extraterrestrial objects continue today, as evidenced by Tunguska (1908; e.g., Farinella *et al.* 2001; Kvasnytsya *et al.* 2013; Robertson and Mathias, 2019) and Chelyabinsk (2013; Popova *et al.* 2013; Artemieva and Shuvalov, 2016). Cometary dust particles are regularly seen during meteor showers and have been collected and studied (e.g., Wild2, Stardust Mission; Brownlee *et al.* 2006), and orbiting spacecraft and ground-based observations provide evidence for current impacts on the Moon (Speyerer *et al.* 2016; Madiedo *et al.* 2019) and Mars (Beyer 2019; Dauber *et al*. 2019). Modern-day accretion rates of ~$10^5$ kg/yr (e.g., Anders 1989) have now been revised upwards to ~$10^7$ kg/yr (Love and Brownlee 1993). Thus, GLA and other molecules are likely continually delivered. It is impossible, though, to predict under what conditions the availability of these molecules was/is preserved or enhanced, so we leave it to the reader to adjust the values in Tables 2, 3, and 4 according to their preferred model of survivability. Even with the assumption that almost all (~99%) of the organic molecules are destroyed during ablation, though, 1% of the value of organic molecules (see Table 4) remained to be delivered to an early Earth.

### 4.1 GLA and Pre-biotic Chemistry

The formose reaction may have played an important role in the prebiotic synthesis of ribose and other larger sugars (Breslow, 1959) and recent work has focused on the role that minerals might have played in controlling the synthetic outcomes of this reaction. Most studies start with a mixture of GLA and formaldehyde because the first step of the formose reaction (dimerization of formaldehyde to form glycolaldehyde) is sluggish and only proceeds with appreciable yields at high pHs. Nisbet and Sleep (2001) have suggested that the oceans on the Archean Earth were acidic, which may have had a negative effect on the rates and yields of this reaction. Perez-Jimenez *et al*. (2011), however, have shown that acidic conditions may have been required for the evolution of biotic enzymes. A study of Precambrian thioredoxin enzymes (Trx) provides evidence that they evolved over ~3 billion years, adapting from hotter, more acidic conditions to today's cooler, less acidic environment (Halevy and Bachan, 2017). Under alkaline conditions, the mixture of glycolaldehyde and formaldehyde in the presence of borate would also create biomolecules (e.g., pentoses; Neveu *et al.* 2013). Whatever the pH of the early Earth, it may not have been global, and a single impact of a GLA-rich comet would have seeded that environment with at least a portion of its GLA. The resulting catalysis of the formose reaction would have led to the synthesis of other biomolecules, including sugar alcohols and sugar acids, all vital to life as we know it — they act as energy sources and are also components of nucleic acids (RNA, DNA) and cell membranes.

The increased availability of GLA through cometary delivery could also have played an important role in the synthesis of simple amino acids. Early reports of the reactions of aldehydes with hydrogen cyanide (the Strecker Synthesis, e.g., Kendall and McKenzie, 1929; Clarke and Bean, 1931) provided for an abiotic route for the synthesis of α-amino acids. More recently, Kebukawa *et al.* (2017) have been able to create α-, β-, and γ-amino acids up to five carbons by mixing formaldehyde, GLA, and ammonia in the presence of liquid water and macromolecular organic solids similar to the chondritic insoluble organic matter (that may be linked to comets; Cody *et al*. 2011). These synthetic pathways, along with the vast diversity of amino acids that have been found in a large number of meteorites (e.g., Burton *et al*. 2011; Cody *et al*. 2011; Glavin *et al.* 2012a, Glavin *et al*. 2012b, Koga and Hiroshi 2017, Pizzarello *et al.* 2008, Pizzarello *et al*. 2012) provide a rich reservoir of organic molecules available for prebiotic chemistry.

## 4.2 Implications for the Origin of Life

Evidence from terrestrial zircons (e.g., Wilde *et al.* 2001) indicates that water and land masses existed around 4.2-4.1 Ga, implying a cool early Earth that was not as "Hadean" as previous studies (e.g., Abramov *et al.* 2013) have suggested. Thus, the effects of impactors may not have been severe enough to cause "impact sterilization" (e.g., Maher and Stevenson, 1988; Sleep *et al.* 2001; Zahnle and Sleep, 1997; Nisbet and Sleep, 2001) or destruction of biomolecules (e.g., Chyba *et al.* 1990). Indeed, recent evidence from dynamic modeling, lunar and other samples, and orbital data shows that the Moon's bombardment rate (and by proxy, the Earth's) may have been drawn out and not especially cataclysmic (e.g., Zellner 2017). It is known that biomolecules, including the ones described here, could exist in temperate conditions. Life very early in Earth's history has been suggested by Djokic *et al.* (2017), who presented evidence for the earliest signs of life on land at 3.48 Ga; by Nutman *et al.* (2016), who posited microbial fossils at 3.7 Ga; by Schidlowski (1988) and Rosing (1999), who reported biogenic C isotopes at ~3.8 Ga; and by Bell *et al.* (2015), who reported putative biogenic carbon at 4.1 Ga.

## 4.3 Other Planetary Bodies

As an inner solar system body, Mars also experienced impact by comets, which may have delivered as much as $10^{20}$ kg of material (e.g., Rickman *et al.* 2017 Morbidelli *et al.* 2018), assuming a monotonic decline (e.g., Werner 2019; Moser *et al.* 2019). Rickman *et al.* (2016) estimated that, based on the size of Mars, as well as the transfer efficiency into the inner solar system for the population of cometary impactors in that study, the size (radius) of comets impacting Mars would range from ~40 km to ~140 km; the lower size limit is approximately three times that of the minimum radius suggested for Hale-Bopp, meaning that even larger amounts of organic molecules could survive delivery or be produced during impact. Evidence shows that Mars was once warmer and wetter than it is now (Grotzinger *et al.* 2015 and references therein) and the results reported herein strengthen the case that Mars may possess more biologically important molecules than those already detected (e.g., reduced organic compounds in the form of macromolecules, Eigenbrode *et al.* 2018; amino acids of martian origin, Callahan *et al.* 2013).

Comets, because they originate in the outer solar system, are most likely to have dominated the impactor populations in the outer solar system, especially in the first billion years of solar system history. This is an especially intriguing scenario for Europa and Enceladus, icy moons of Jupiter and Saturn, respectively, on which water has been detected and which may have conditions suitable for life. Though the timing of comet delivery to planetary bodies in the outer solar system is somewhat unconstrained (e.g., Jørgensen *et al.* 2009; Nesvorný *et al.* 2017; Rickman *et al.* 2017; Morbidelli *et al.* 2018), studies of current impact rates in the outer solar system indicate generally high impact rates, especially for comets whose diameters are larger than a few kilometers (e.g., Zahnle *et al.* 2003). For example, Europa is likely to experience an impact by a 2-km-diameter impactor every 5- or 6 million years (Zahnle *et al.* 2003) and was likely to experience more impacts earlier in its history, making the upcoming search for prebiotically relevant molecules on this icy body in particular extremely compelling (e.g., Blaney *et al.* 2019).

## 5. Conclusions

Comets were likely abundant in the early solar system and likely impacted young planets and moons, delivering organic materials necessary for creating the building blocks of life as we know it. Applying results of experiments that determined survival and production yields of sugar molecules to known values of observed GLA in a variety of comets showed that up to $10^{23}$ kg of GLA and other sugar molecules would have been delivered to or produced during impact onto a planetary body during its first billion years. Regardless of the size or origin of the comet, results reported herein show that even with a high degree of uncertainty in comet diameters and volumes, the amounts of GLA, EG and simple sugars like threose that could have been delivered to an early Earth are quite large. These results provide evidence that large amounts of GLA and more complex impact-produced sugar molecules may have been readily available on habitable moons or planets, especially during the era of late heavy bombardment (~4.2 to ~3.7 billion years ago). Furthermore, the presence and availability of these biomolecules under the right conditions may be important for understanding the origin of life as we know it and may have driven pre-biotic chemical reactions that lead to, for example, ribose, the five-carbon sugar in RNA.

**Acknowledgments**

The authors acknowledge funding from the NASA Exobiology and Evolutionary Biology Program (10-EXO10-0109) and Albion College's Foundation for Undergraduate Research, Scholarship and Creative Activity. Funding for conference travel to present results of this study was provided, in part, by a grant from the Hewlett-Mellon Fund for Faculty Development at Albion College, Albion. MI. We thank two anonymous reviewers for their thoughtful comments and suggestions that helped to improve this manuscript. We also thank Mark Cintala, Frank Cardenas, and Roland Montes at the NASA Johnson Space Center's Experimental Impact Facility for their assistance during the impact experiments; David Carey at Albion College for assistance in the chemistry labs; and John Delano for providing the montmorillonite clays.

**Author Disclosure Statement**

No competing financial interests exist.